\documentclass[dvipsnames,format=sigconf,anonymous=False,review=False]{acmart}
\usepackage{amsmath} 
\usepackage{float}
\usepackage{algorithm}
\usepackage{algpseudocode}
\usepackage{array}
\usepackage{subcaption} % Use subcaption instead of subfigure
\usepackage{multirow}

\usepackage{xcolor,colortbl}
\usepackage{hyperref} % For hyperlinks in references
\usepackage[utf8]{inputenc}
\usepackage{tikz}
\usepackage{ulem}

\setcopyright{acmlicensed}
\copyrightyear{2025}
\acmYear{2025}
\acmDOI{XXXXXXX.XXXXXXX}
\acmConference[GECCO '25]{The Genetic and Evolutionary Computation Conference}{July 14--18,
  2025}{M\'alaga, Spain}
\acmISBN{978-1-4503-XXXX-X/18/06}

\begin{document}

% Define a custom very light gray color
\definecolor{verylightgray}{gray}{0.90}
\definecolor{Col0}{HTML}{1f77b4} % DMRG 2
\definecolor{Col1}{HTML}{2ca02c} % DMRG 0.10
\definecolor{Col2}{HTML}{ff7f0e} % DMRG 0.20
\definecolor{Col3}{HTML}{9467bd} % GNN 9467bd
\definecolor{Col4}{HTML}{d62728} % GA-OC d62728
\definecolor{Col5}{HTML}{8c564b} % GA 500
\definecolor{Col6}{HTML}{e377c2} % GA 1000
\definecolor{Col7}{HTML}{7f7f7f} % GA 2000


% \title{Performance Comparison of Classical and Quantum-Inspired Algorithms for Combinatorial Optimization Problems}
\title{Comparative Analysis of Classical and Quantum-Inspired Solvers: A Preliminary Study on the Weighted Max-Cut Problem}

% \title{Evaluating Classical and Quantum-Inspired Algorithms for the Max-Cut Problem}


\author{Aitor Morais}
\orcid{https://orcid.org/0009-0003-4148-8874}
\affiliation{%
  \institution{University of Deusto (D4K)}
  \city{Bilbao}
  \country{Spain}
}
\email{aitor.morais@deusto.es}

\author{Eneko Osaba}
\affiliation{%
  \institution{Tecnalia, Basque Research and \\ Technology Alliance (BRTA)}
  \city{Derio}
  \country{Spain}}
\email{eneko.osaba@tecnalia.com}
\orcid{https://orcid.org/0000-0002-3950-1668}

\author{Iker Pastor}
\orcid{https://orcid.org/0000-0002-3068-6248}
\affiliation{%
  \institution{University of Deusto (D4K)}
  \city{Bilbao}
  \country{Spain}
}
\email{iker.pastor@deusto.es}

\author{Izaskun Oregi}
\affiliation{%
  \institution{Tecnalia, Basque Research and \\ Technology Alliance (BRTA)}
  \city{Derio}
  \country{Spain}}
\email{izaskun.oregui@tecnalia.com}
\orcid{https://orcid.org/0000-0002-3950-1668}

\renewcommand{\shortauthors}{Morais et al.}

\begin{abstract}
Combinatorial optimization is essential across numerous disciplines. Traditional metaheuristics excel at exploring complex solution spaces efficiently, yet they often struggle with scalability. Deep learning has become a viable alternative for quickly generating high-quality solutions, particularly when metaheuristics underperform. In recent years, quantum-inspired approaches such as tensor networks have shown promise in addressing these challenges. Despite these advancements, a thorough comparison of the different paradigms is missing. This study evaluates eight algorithms on Weighted Max-Cut graphs ranging from 10 to 250 nodes. Specifically, we compare a Genetic Algorithm representing metaheuristics, a Graph Neural Network for deep learning, and the Density Matrix Renormalization Group as a tensor network approach. Our analysis focuses on solution quality and computational efficiency (i.e., time and memory usage). Numerical results show that the Genetic Algorithm achieves near-optimal results for small graphs, although its computation time grows significantly with problem size. The Graph Neural Network offers a balanced solution for medium-sized instances with low memory demands and rapid inference, yet it exhibits more significant variability on larger graphs. Meanwhile, the Tensor Network approach consistently yields high approximation ratios and efficient execution on larger graphs, albeit with increased memory consumption.
\end{abstract}


\maketitle
\section{Introduction}

Combinatorial Optimization (CO) problems focus on finding the optimal solution --or solutions for multimodal problems--, from a set of feasible options. These problems are highly relevant across various real-world applications where optimal decision-making is essential. For instance, the Traveling Salesman Problem \cite{TSP} is crucial for optimizing delivery routes and reducing transportation costs. Similarly, the Job Shop Scheduling problem \cite{jain1999deterministic} minimizes processing time in manufacturing. The Max-Cut problem \cite{trevisan2009max} also finds applications in social network analysis \cite{fortunato2010community} and image segmentation \cite{yi2012image}.

Metaheuristics \cite{ezugwu2021metaheuristics}, including Genetic Algorithms (GA, \cite{GA}), Particle Swarm Optimization \cite{PSO}, and Ant Colony Optimization (ACO) \cite{dorigo2006ant} have been traditionally employed to explore combinatorial search spaces efficiently. However, they may not guarantee optimality and can become trapped in local minima \cite{rajwar2023exhaustive}. Hence, Machine learning techniques have emerged as promising alternatives for solving combinatorial optimization problems \cite{bengio2021machine}. Deep learning approaches such as reinforcement learning, attention mechanisms, and imitation learning have also been explored to replace handcrafted heuristics in decision-making processes. In particular, Graph Neural Networks (GNNs) \cite{cappart2023combinatorial, khalil2017learning} have demonstrated the ability to capture structural relationships in CO problems. Recently, Schuetz et al. \cite{schuetz2022combinatorial} employed a physics-informed GNN-based architecture with a differentiable loss function and a projection step to address the Max-Cut problem.

Building on novel computational paradigms, quantum computing introduces algorithms that leverage the unique properties of quantum mechanics to process information in ways that classical computers cannot. These quantum features enable quantum algorithms to explore and solve complex problems more efficiently than classical algorithms, such as the Quantum Approximate Optimization Algorithm (QAOA, \cite{QAOA}), Variational Quantum Eigensolver (VQE, \cite{VQE}), and quantum annealing (QA, \cite{qannealing}) that leverage quantum features to process information. However, current quantum hardware characterized by the Noisy Intermediate Scale Quantum (NISQ \cite{NISQ}) era limits their real-world applicability \cite{djidjev2018efficient,guerreschi2019qaoa}. This challenge has spurred the development of quantum-inspired algorithms based on Tensor Networks (TN, \cite{orus2019tensor}). Recent works \cite{hao2022quantum, nakada2024quick} have successfully mapped CO problems to the quantum domain by encoding classical objectives into quantum Hamiltonians and employing techniques such as imaginary time evolution. Although there is significant theoretical progress in this line, as far as we know, there is a lack of a complete comparative study evaluating the performance of these innovative approaches against traditional algorithms. 

The core of our work involves benchmarking three distinct solvers, each representing a unique computational paradigm. We utilize a GA as a classical metaheuristic, a GNN as a machine learning-based solver, and a TN approach combined with the Density Matrix Renormalization Group (DMRG, \cite{catarina2023density}) from the quantum-inspired framework. To ensure a thorough comparison, we employ a systematic procedure that includes conventional statistical tests to evaluate the performance of these solvers. We apply the algorithms to a set of the Weighted Max-Cut problem instances in terms of the quality of the results and computational efficiency. The results enable us to identify the strengths and limitations of each paradigm and provide insights into their suitability for different problem scales and resource constraints.

The paper is organized as follows: Section \ref{sec2:sota} reviews state of the art and identifies gaps in current research; Section \ref{sec2:background} provides background on TNs; Section \ref{sec3:algorithms} details the algorithms and their mathematical formulations; Section \ref{sec4:experimentation} describes the experimental setup; Section \ref{sec5:results} discusses the obtained results, and Section \ref{sec6:conclusions} concludes with directions for future research.

\section{Literature Review}
\label{sec2:sota}
The literature on benchmarking classical metaheuristics for combinatorial optimization is extensive, with foundational works such as \textit{Talbi’s Metaheuristics: From Design to Implementation} \cite{talbi2009metaheuristics} and\textit{ Blum and Roli’s comprehensive survey on metaheuristics} \cite{blum2003metaheuristics} providing critical insights. This section offers a concise overview to contextualize the current study, focusing on metaheuristics, deep learning, and TN methods without aiming for exhaustive coverage.

\paragraph{Metaheuristics}
Wang et al. \cite{hui2012comparison} compared a GA, Hopfield neural network, and ACO on the TSP, evaluating efficiency, solution quality, time complexity, space complexity, and implementation difficulty. Results showed GAs produced high-quality solutions with specialized operators but required careful tuning to avoid suboptimal convergence, while ACO struggled with local optima without effective pheromone resetting, and Hopfield networks faced challenges due to intensive matrix updates. GAs were computationally expensive but effective, often nearing optimal solutions.

In addressing the fuzzy Max-Cut problem, Wang et al. \cite{wang2010maximum} developed a hybrid GA with fuzzy simulation, formulating models based on credibility criteria. Their approach yielded near-optimal solutions but encountered scalability issues due to computational demands. Similarly, Soares et al. \cite{soares2023genetic} enhanced GAs for Max-Cut with optimality cuts, improving performance, yet lacked comparisons with deep learning and TN methods, highlighting a research gap.

\paragraph{Deep Learning}
Classical metaheuristics like GAs, ACO, and Hopfield networks operate in polynomial time per iteration but require multiple iterations, with GAs benefiting from parallelism (e.g., GPU use) yet facing scalability challenges for large instances. Deep learning approaches, such as Bello et al.’s \cite{bello2016neural} pointer network for TSP and Yao et al.’s \cite{yao2019experimental} GNNs for Max-Cut, show promise but lack rigorous benchmarking against classical heuristics. Li et al. \cite{li2022rethinking} and Schuetz et al. \cite{schuetz2022combinatorial} improved GNN performance with problem-specific adaptations, while Nath and Kuhnle \cite{nath2024benchmark} introduced MaxCut-Bench to standardize comparisons, revealing GNNs often underperform baseline heuristics without tailoring.

Stoudenmire and Schwab \cite{stoudenmire2016supervised} introduced TN-based supervised learning for efficient, interpretable optimization, excelling in tasks like MNIST classification. Quantum-inspired methods, as in the Open-Pit Mining Problem \cite{young1988introduction}, use TNs to encode constraints and find optimal solutions, maintaining accuracy even with reduced bond dimensions, though runtime and resource details are absent.

\paragraph{Tensor Networks}

Gardiner et al. \cite{gardiner2024tensor} linked TNs to Estimation of Distribution Algorithms, finding that simpler models with added noise can enhance solution quality, outperforming classical GAs and other TN approaches on benchmarks like Knapsack and Max-3SAT. However, they omit runtime and complexity analyses, focusing solely on solution quality.

\paragraph{Motivation} With the development of quantum-inspired methods, such as TN-based algorithms, it is necessary to expand comparisons to include all three paradigms. Nath et al. \cite{nath2024benchmark}, for instance, have recently compared traditional heuristics and GNN-based methods, omitting quantum-inspired methods. This study addresses this gap through empirical comparisons across diverse problem instances. Specifically, we include a GA, a GNN, and a TN-based scheme inspired by quantum physics. As a preliminary study, we implement clean and minimal versions of each method.

We utilize GAs -- including problem-specific variants ~\cite{soares2023genetic} -- due to their extensive empirical validation in combinatorial optimization literature. Regarding deep learning approaches, we apply GNNs for their effectiveness in modeling combinatorial problems \cite{cappart2023combinatorial, schuetz2022combinatorial} and used a TN-based approach with the DMRG algorithm for the quantum-inspired paradigm. We include a detailed description of compared algorithms in Section \ref{sec3:algorithms}.

Our benchmarking framework assesses solution quality and computational performance, including execution time and memory usage. In line with the systematic benchmarking approach proposed by Lorenz et al. \cite{lorenz2025systematic}, our work constitutes an application-level benchmark, with metrics aligned to established recommendations in the literature. The insights from this study provide a foundation for future work exploring hybrid optimization approaches that combine the strengths of metaheuristics, deep learning, and TNs.

\section{Tensor Network Basics}\label{sec2:background}

In this section, we provide an overview of the fundamental concepts of TN. We begin by introducing the MPS then we describe the MPO, and we finish by reviewing the DMRG optimization algorithm. These concepts are essential for understanding the quantum-inspired approach we use in the experimental part of this paper. For a more comprehensive introduction to these tensor network objects, we refer interested readers to \cite{nut, catarina2023density,schollwock2011density}.

Before delving into TN objects, it is essential to introduce Dirac notation (or bra-ket notation), the standard formalism used in quantum mechanics for representing quantum states as vectors in a Hilbert space. In this notation, a ket $|\psi\rangle$ represents the quantum state $\psi$. In a two-level quantum system (e.g., a single qubit state), the state $|\psi\rangle$ can be expressed as a superposition of the basis states $|0\rangle$ and $|1\rangle$ as $$|\psi\rangle = \alpha |0\rangle + \beta |1\rangle,$$ where $\alpha$ and $\beta$ are complex coefficients satisfying normalization condition $|\alpha|^2+|\beta|^2=1$. Here, the vectors $|0\rangle=\left[1,0\right]^T$ and $|1\rangle=\left[0, 1\right]^T$ represent the computational basis, representing the classical binary states $0$ and $1$, respectively. Additionally, the bra  $\langle\psi|$, corresponds to the Hermitian conjugate (complex conjugate transpose) of the ket $|\psi\rangle$.

\subsection{Matrix Product States}\label{subsec:MPS}
Dirac notation provides a compact way to represent more complex systems. For systems with $N$ qubits, the state $|\psi\rangle$ resides in a $2^N$-dimensional space. Note that the number of elements required to represent this vector scales exponentially with the number of qubits. This rapid growth makes quantum states computationally difficult to store and manipulate using classical computers. To address this issue, MPS emerged as a powerful tool in quantum physics. By breaking down a quantum state $|\psi\rangle$ into a linear set of tensors, MPS enhances scalability by assuming a low error \cite{orus2019tensor,verstraete2006matrix}.

\begin{figure}[t]
    \centering
    \begin{subfigure}{0.45\linewidth}
        \centering
        \includegraphics[scale=0.5]{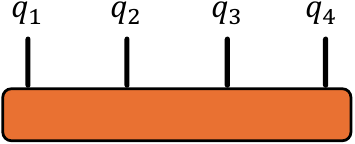}
        \caption{}
        \label{subfig:a}
    \end{subfigure}
    \begin{subfigure}{0.45\linewidth}
        \centering
        \includegraphics[scale=0.5]{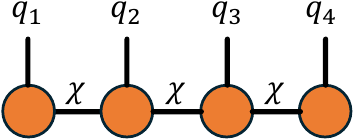}
        \caption{}
        \label{subfig:b}
    \end{subfigure}
    
    \caption{Diagrammatic representation of (a) the $c_{q_1,q_2,...,q_N}$ tensor and (b) its MPS representation for a system of $N=4$ qubits. Here, filled shapes are used to represent tensors, while lines extending from these shapes denote their indices. }
    \label{fig:mps}
\end{figure}

Consider a quantum state $|\psi\rangle$ of $N$ qubits, expressed as

\begin{equation}
    |\psi\rangle = \sum_{q_1, \ldots, q_N}c_{q_1,...,q_N} |q_1\cdots q_N\rangle,
\end{equation}

where $q_i\in\{0,1\}$ denotes the state of the $i$-th qubit, which can be either $|0\rangle$ or $|1\rangle$. The term $|q_1\cdots q_N\rangle$ represents a tensor product of the basis states of individual qubits, and $c_{q_1,...,q_N}$ are the complex-valued coefficients associated with each combination of the basis states\footnote{For example, in a 4 qubit system $c_{0011}$ gives the amplitude of the system to be in the $|0011\rangle$ basis state.}. Then, the $N$-site MPS decomposes the tensor $c_{q_1,...,q_N}$ into a sequence of local tensors $A^{q_i}$ as follows:
\begin{equation}\label{eq:mps}
    c_{q_1,q_2,...,q_N} = \sum_{\chi_1,\dots,\chi_N} A^{q_1}_{\chi_1} A^{q_2}_{\chi_1,\chi_2} \dots A^{q_N}_{\chi_{N-1}}.
\end{equation}
Here, the intermediate indices $\chi_1, \ldots,\chi_{N-1}$, called bond dimensions, control the extend of correlations (more precisely entanglement\footnote{In quantum mechanics entanglement is a type of correlation between quantum systems where the state of one qubit depends on others.}) that can be captured between different qubits, larger bond dimensions allow for higher entanglement but increase computational cost. In this paper, we assume the bond-dimension is $\chi_1=\cdots=\chi_{N-1}$, and we use $\chi$ to denote it. An example schematic description of an MPS for a 4-qubit system 
is depicted in Figure~\ref{fig:mps}.

\subsection{Matrix Product Operators}\label{subsec:MPO}

Having established the MPS scheme, we now turn to the question of how to represent operators acting on these states. In quantum mechanics, operators describe interactions or measurable quantities of a system, such as the Hamiltonian, which encodes the energy of the system. Mathematically, in an $N$-qubit system, the Hamiltonian operator is a $2^N\times2^N$ Hermitian matrix. Therefore, as with quantum states, storing and manipulating operators in their full matrix form quickly becomes infeasible for large systems. This motivates the use of MPO, a tensor-based representation analogous to MPS. Formally, an operator $\hat{H}$ acting on a $N$-qubit system is given by
\begin{equation}
\hat{H}=\sum_{\substack{q_1,\dots,q_N \\ q'_1,\dots,q'_N}} c_{q_1,...,q_N}^{q'_1,...,q'_N} \left| q_1,\dots,q_N \right> \left< q'_1\dots,q'_N \right|,
\end{equation}
where $c_{q_1,...,q_N}^{q'_1,...,q'_N}$ represent the complex-valued coefficients of the operator in the computational basis. Note that these coefficients form a $2N$-rank tensor, which can be intractable for large $N$. Therefore, instead of storing the tensor explicitly, MPO expresses it as a product of smaller tensors,
\begin{equation}\label{eq:mpo}
c_{q_1,...,q_N}^{q'_1,...,q'_N}=\sum_{\alpha_1, ..., \alpha_{N-1}}W_{\alpha_1}^{q_1q'_1}W_{\alpha_1\alpha_2}^{q_2q'_2}\cdots W_{\alpha_{N-1}}^{q_Nq'_N},
\end{equation}
where $W_{\chi_{i-1}\chi_i}^{q_iq'_i}$ are local tensors associated with each qubit. 

\begin{figure}[t]
    \centering
    \begin{subfigure}{0.45\linewidth}
        \centering
        \includegraphics[scale=0.5]{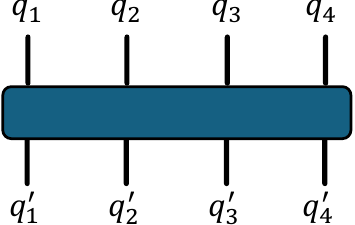}
        \caption{}
        \label{subfig:a}
    \end{subfigure}
    \begin{subfigure}{0.45\linewidth}
        \centering
        \includegraphics[scale=0.5]{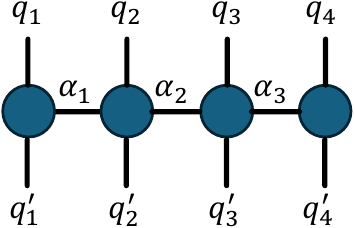}
        \caption{}
        \label{subfig:b}
    \end{subfigure}
    
    \caption{Diagrammatic representation of (a) the $c_{q_1,...,q_N}^{q'_1,...,q'_N}$ tensor, and (b) its MPO representation for a system of $N=4$ qubits.}
    \label{fig:mpo}
\end{figure}

A schematic description of an MPO for a 4-qubit system 
is shown in Figure~\ref{fig:mpo}.

\subsection{Densitity Matrix Renormalization Group }\label{subsec:DMRG}

Given a Hamiltonian operator represented as an MPO and corresponding quantum states as an MPS, the variational optimization algorithm DMRG finds the system's minimum energy by iteratively minimizing the expectation value of the Hamiltonian. 

Let $\hat{H}$ be the Hamiltonian of a $N$-qubit system, and let $|\psi\rangle$ represent a state of the system. The DMRG tackles the optimization problem given by
\begin{equation}
    \min_{|\psi\rangle} E\left(|\psi\rangle\right),
\end{equation}
where the energy function is defined as
\begin{equation}
E(|\psi\rangle)=\frac{\langle \psi|H|\psi\rangle}{\langle\psi|\psi\rangle}.
\end{equation}
In doing so, the algorithm uses Eq. \eqref{eq:mpo} and \eqref{eq:mps} to compute the expected value. However, instead of contracting the entire TN to evaluate the energy and improve the initial guess of $|\psi\rangle$ (which is usually random). Iteratively optimizes local tensors $A^{q_i}$ of the MPS by solving the eigenvalue problem more efficiently.

To apply this process 
to classical CO problems, the objective function must be encoded as a Hamiltonian, and the feasible solution space must be represented as a set of quantum states.

\section{Description of Compared Algorithms}\label{sec3:algorithms}

We begin this section by presenting the mathematical formulation of the Weighted Max-Cut problem, a fundamental combinatorial optimization problem that serves as the focus of our experimentation. Following this, we provide a detailed description of the three architectures being compared: the GA, the GNN, and the DMRG.

\subsection{The Weighted Max-Cut Problem}\label{subsec:MAXCUT}
Consider the undirected graph $G = (\mathcal{V}, E)$, where $\mathcal{V}$ is the set of vertices with $|\mathcal{V}| = N$, and $E$ is the set of edges, and $w_{ij} \ge 0$ represents the weight associated with the edge $(i, j) \in E$, satisfying $w_{ij}=w_{ji}$. The goal of the Max-Cut problem is to split the vertex set $\mathcal{V}$ into two disjoint subsets, $\mathcal{S}$ and $\mathcal{T}$, such that the total weight of the edges crossing between the two sets is maximized. 

To mathematically represent the problem, we define a binary vector $\mathbf{x}\in\{0,1\}^N$, where each element $x_i$ indicates the set membership of the node $i$. That is, $x_i=0$ if $i\in\mathcal{S}$ and $x_i=1$ if $i\in\mathcal{T}$. Using this representation, the Weighted Max-Cut objective function can be written as
\begin{equation}\label{eq:max-cut1}
    \text{Max-Cut}(\mathbf{x})=\sum_{i<j}w_{ij}\left[x_i+x_j-2x_ix_j\right].
\end{equation}
Note that this formulation captures the contribution of an edge $(i,j)$ to the cut only when nodes $i$ and $j$ belong to different subsets, i.e., $x_i\neq x_j$.

Note that Eq. \eqref{eq:max-cut1} leads to the Quadratic Unconstrained Binary Optimization (QUBO, \cite{glover2018tutorial}) formulation, where the objective function represents a binary optimization problem with a quadratic loss function and no constraints. The general form of a QUBO problem is given by
\begin{equation}\label{eq:qubo_formulation}
    H_{\text{QUBO}}=\mathbf{x}^TQ\mathbf{x},
\end{equation}
where $Q$ is an $N\times N$ symmetric matrix representing the quadratic objective function. The QUBO model is central to quantum optimization because it maps naturally onto the physical systems underlying quantum devices. For example, quantum annealers like those developed by D-Wave \cite{neukart2017traffic} solve QUBO problems by finding the ground state of a corresponding quantum system. Moreover, as we will see, this formulation is also interesting for quantum-inspired algorithms.

In the case of the Weighted Max-Cut problem, the QUBO matrix is given by
\begin{equation}\label{eq:qubo_matrix_maxcut}
    Q_{ij} = \begin{cases}
        -w_{ij} & \text{if } i \neq j \\
        \sum_{k\neq i} 2w_{ik} & \text{if } i=j
    \end{cases},
\end{equation} where we have changed the sign of the objective function to align the ground state search with the maximization criterion of the Weighted Max-Cut.

\subsection{Algorithms}\label{subsec:algorithms}
\subsubsection*{Genetic Algorithm.} A canonical Genetic Algorithm (cGA) has been implemented to address the Weighted Max-Cut problem. Solutions are represented as binary chromosomes $\mathbf{x} \in {0,1}^N$, where each gene $x_i$ indicates the partition assignment for node $i$. The algorithm initializes a population of $P$ individuals with uniformly random bit values. Fitness evaluation employs Eq.~\eqref{eq:max-cut1}, while selection follows a roulette wheel scheme weighted by fitness scores. Genetic operators include single-point crossover with randomized split positions and bit-flip mutation with probability $p_m = 0.1$. An elitism mechanism preserves the top $2\%$ solutions per generation. The termination criterion is set at 1000 generations, ensuring convergence toward optimal Max-Cut solutions while maintaining population diversity through stochastic variation operators. We also implement a problem-specific Max-Cut GA (GA-OC) framework introduced in \cite{soares2023genetic}; this approach incorporates a greedy refinement procedure that locally evaluates node contributions and dynamically adjusts partition assignments to seek optimal configurations. Although the paper states that they use time as the stopping criterion, specifically 1800 seconds, we have employed 1000 iterations as the stopping criterion to ensure fair comparison with our cGA and DMRG methods. It is worth noting that their code was written in C++, whereas ours was implemented in Python. The population and other hyperparameter settings are those specified in the mentioned paper. Specifically, a uniformly random population of 300 individuals is initialized, and 50 new individuals are generated in each iteration. For the tournament, 4 individuals are randomly selected, and the best of them is returned. The crossover method is uniform with two individuals. If the vertex value of the parents is different, the child has a chance of inheriting from the fittsest parent, this procedure generates only one child. Mutation is applied to $20\%$ of the population, excluding the best individual. Each gene has a $10\%$ probability of being mutated.

\subsubsection*{GNN Algorithm} To address the Weighted Max-Cut problem from the machine learning perspective, we implement the GNN algorithm in \cite{schuetz2022combinatorial}, where the authors propose a method that applies GNNs to solve NP-hard problems such as the Max-Cut problem. We define the architecture and training procedure, where the model hyper-parameters are configured according to the authors’ guidelines.

The GNN architecture they propose consists of a two-layer Graph Convolutional layer (GCN), where the first layer takes node embeddings of dimension $d_0=369$ and outputs a representation of dimension $d_1=5$. This layer is followed by a component-wise, non-linear ReLU transformation. Subsequently, the second GCN layer takes this intermediate representation and generates the output layer of size $d_2=1$. Finally, this output is passed through a component-wise sigmoid transformation to yield a soft probability $p_i \in [0, 1]$ for all node $i \in \mathcal{V}$.

The training strategy consists of relaxing the problem Hamiltonian introduced in Eq. \eqref{eq:qubo_formulation} to create a differentiable loss $\mathcal{L}(\theta)$ where the gradient descent method is viable. In doing so, the authors transform the binary decision vector $\mathbf{x}\in\{0,1\}^N$ into a continuous parametrized probability array $\mathbf{p}\left(\theta\right)\in\left[ 0,1\right]^N$ so the Hamiltonian is transformed to
\begin{equation}
    \mathcal{L}(\theta)=\mathbf{p}^T(\theta)Q\mathbf{p}(\theta),
\end{equation} which is differentiable with respect to $\theta$, the parameters of the GNN model. All models are optimized using Adam \cite{kingma2014adam} with a learning rate $l_r=0.00467$.

\subsubsection*{Quantum-Inspired Algorithm} The quantum-inspired framework implemented in this paper consists of applying the DMRG algorithm introduced in Section \ref{subsec:DMRG}. As mentioned before, the goal of the DMRG is to find the MPS that minimizes the energy of an operator represented as an MPO. 

The MPS is initialized randomly from a uniform distribution $U(0,1)$ for a specific bond-dimension $\chi$.

To define the MPO, first, we need to formulate the Max-Cut objective function in Eq. \ref{eq:max-cut1} into a Hamiltonian operator. This is done by mapping the binary variable $x_i$ to a spin variable $z_i$ using the relation $x_i=(1-z_i)/2$, where $z_i\in\{-1, 1\}$. Subsequently, we translate the reformulated objective function into the Hamiltonian operator by replacing spin variables with Pauli-Z matrices $Z_i$. Here, the subscript $i$ indicates that the Pauli-Z matrix is acting on the $i$-th qubit.

As a result, the resulting operator for the Weighted Max-Cut problem is given by, 
\begin{equation}
    \hat{H}=\sum_{i<j} -\frac{w_{ij}}{2} (\mathbb{I}-Z_iZ_j)
\end{equation} where $\mathbb{I}$ is the $2^N\times2^N$ identity matrix, and the $Z_iZ_j$ term represents the tensor product of $N$ terms with $Z_i$ and $Z_j$ positioned in the $i$-th and $j$-th positions, respectively.

Once the Hamiltonian is defined, the next step is to create the MPO. To this end, we rely on the weighted finite automata approach described in \cite{automata}. The process of generating the MPO involves three main steps. First, identify the pattern of the operator using a Weighted finite automaton. Next, transform the pattern to a matrix product diagram, which visually represents the operator's structure. Finally, the MPO matrices are generated based on the diagram. While the described process may seem relatively straightforward, it is important to highlight that the complexity arises when implementing long-range interactions. For this work, we focused on developing an algorithm capable of generalizing any Weighted Max-Cut problem. Since this problem involves long-range interactions rather than nearest-neighbor ones, the creation of the automata is non-trivial and requires significant time: first, to understand how automata operate, and then to represent it effectively. Additionally, we aimed to generalize an algorithm that can construct a TN for all possible scenarios of the Weighted Max-Cut problem. Crucially, the symmetry of the Weighted Max-Cut problem allowed us to design a TN with only $N-1$ tensors (where $N$ is the number of nodes ) with maximum bond dimension \(\chi=N \). Beyond this, we minimized the number of automata states to reduce the bond dimension of the resulting MPO. Simplifying computations, which accelerates DMRG calculations by reducing the number of operations required. It is worth noting that our MPO is an exact representation of the Hamiltonian of the problem and not an approximation of the Hamiltonian. 
\begin{table*}[htbp]
    \centering
    \caption{Average approximation ratio and standard deviation for DMRG, GNN, GA-OC, and cGA.}
    \label{tab:energy_tab_reduced}
    \resizebox{\textwidth}{!}{%
    \begin{tabular}{c||ccc|c|c|ccc}
    \toprule
    \multirow{2}{*}{$N$}
      & \multicolumn{3}{c|}{DMRG}
      & \multirow{2}{*}{GNN}
      & \multirow{2}{*}{GA-OC}
      & \multicolumn{3}{c}{cGA} 
      \\
    & $\chi_{\text{bond}}=2$
      & $\chi_{\text{bond}}=0.10N$
      & $\chi_{\text{bond}}=0.20N$
      &
      & %\texttt{Pop.Size}=300
      & \texttt{Pop.Size}=500
      & \texttt{Pop.Size}=1000
      & \texttt{Pop.Size}=2000
      \\
    \midrule
    \rowcolor{verylightgray}
    10
    & (0.996, 0.012)
    & (0.976, 0.027)
    & (0.996, 0.012)
    & (0.878, 0.088)
    & \textbf{(1.000, 0.000)}
    & \textbf{(1.000, 0.000)}
    & \textbf{(1.000, 0.000)}
    & \textbf{(1.000, 0.000)}
    \\
    20
    & (0.981, 0.010)
    & (0.981, 0.010)
    & (0.979, 0.000)
    & (0.871, 0.121)
    & (0.979, 0.000)
    & (0.995, 0.009)
    & \textbf{(1.000, 0.000)}
    & \textbf{(1.000, 0.000)}
    \\
    \rowcolor{verylightgray}
    40
    & (0.969, 0.008)
    & (0.962, 0.013)
    & (0.961, 0.014)
    & (0.904, 0.070)
    & (0.977, 0.000)
    & (0.983, 0.008)
    & (0.984, 0.009)
    & \textbf{(0.987, 0.010)}
    \\
    50
    & \textbf{(0.986, 0.012)}
    & (0.973, 0.007)
    & (0.973, 0.007)
    & (0.956, 0.010)
    & (0.978, 0.003)
    & (0.960, 0.009)
    & (0.976, 0.004)
    & (0.976, 0.007)
    \\
    \rowcolor{verylightgray}
    60
    & \textbf{(0.990, 0.005)}
    & (0.986, 0.005)
    & (0.988, 0.009)
    & (0.943, 0.084)
    & (0.965, 0.009)
    & (0.962, 0.007)
    & (0.970, 0.005)
    & (0.980, 0.006)
    \\
    80
    & \textbf{(0.987, 0.006)}
    & (0.981, 0.010)
    & (0.981, 0.007)
    & (0.926, 0.072)
    & (0.968, 0.003)
    & (0.949, 0.008)
    & (0.953, 0.005)
    & (0.960, 0.003)
    \\
    \rowcolor{verylightgray}
    90
    & \textbf{(0.995, 0.004)}
    & (0.980, 0.006)
    & (0.986, 0.006)
    & (0.963, 0.025)
    & (0.977, 0.001)
    & (0.949, 0.002)
    & (0.958, 0.004)
    & (0.963, 0.003)
    \\
    100
    & \textbf{(0.988, 0.006)}
    & (0.982, 0.004)
    & (0.987, 0.003)
    & (0.954, 0.021)
    & (0.969, 0.002)
    & (0.944, 0.003)
    & (0.953, 0.005)
    & (0.955, 0.004)
    \\
    \rowcolor{verylightgray}
    120
    & \textbf{(0.987, 0.001)}
    & (0.984, 0.003)
    & (0.986, 0.004)
    & (0.937, 0.035)
    & (0.976, 0.000)
    & (0.945, 0.005)
    & (0.954, 0.002)
    & (0.953, 0.003)
    \\
    140
    & \textbf{(0.989, 0.004)}
    & (0.983, 0.002)
    & (0.987, 0.004)
    & (0.948, 0.032)
    & (0.977, 0.002)
    & (0.943, 0.003)
    & (0.949, 0.002)
    & (0.950, 0.002)
    \\
    \rowcolor{verylightgray}
    150
    & (0.986, 0.003)
    & \textbf{(0.987, 0.003)}
    & \textbf{(0.987, 0.003)}
    & (0.970, 0.010)
    & (0.977, 0.004)
    & (0.947, 0.005)
    & (0.951, 0.004)
    & (0.951, 0.002)
    \\
    170
    & (0.983, 0.003)
    & (0.985, 0.003)
    & \textbf{(0.986, 0.004)}
    & (0.923, 0.102)
    & (0.972, 0.001)
    & (0.943, 0.003)
    & (0.948, 0.002)
    & (0.948, 0.002)
    \\
    \rowcolor{verylightgray}
    200
    & \textbf{(0.989, 0.001)}
    & (0.985, 0.003)
    & (0.987, 0.003)
    & (0.868, 0.290)
    & (0.978, 0.001)
    & (0.947, 0.001)
    & (0.950, 0.002)
    & (0.951, 0.001)
    \\
    220
    & \textbf{(0.988, 0.002)}
    & \textbf{(0.988, 0.002)}
    & (0.986, 0.002)
    & (0.830, 0.290)
    & (0.982, 0.000)
    & (0.948, 0.001)
    & (0.951, 0.002)
    & (0.952, 0.002)
    \\
    \rowcolor{verylightgray}
    250
    & \textbf{(0.990, 0.003)}
    & (0.989, 0.002)
    & (0.987, 0.002)
    & (0.970, 0.008)
    & (0.984, 0.002)
    & (0.949, 0.002)
    & (0.951, 0.001)
    & (0.953, 0.000)
    \\
    \bottomrule
    \end{tabular}
    }
\end{table*}

\begin{table*}[htbp]
    \centering
    \caption{Execution times in \textbf{minutes} (mean and standard deviation) for DMRG, GNN, GA-OC, and cGA.
             The best (lowest) \emph{average} time for each $N$ is in bold.}
    \label{tab:time_in_minutes_reduced}
    \resizebox{\textwidth}{!}{%
    \begin{tabular}{c||ccc|c|c|ccc}
    \toprule
    \multirow{2}{*}{$N$}
      & \multicolumn{3}{c|}{DMRG}
      & \multirow{2}{*}{GNN}
      & \multirow{2}{*}{GA-OC}
      & \multicolumn{3}{c}{cGA} 
      
      \\
    & $\chi_{\text{bond}}=2$
      & $\chi_{\text{bond}}=0.10N$
      & $\chi_{\text{bond}}=0.20N$
      &
      & %\texttt{Optimal-Cuts GA}
      & \texttt{Pop.Size}=500
      & \texttt{Pop.Size}=1000
      & \texttt{Pop.Size}=2000
      \\
    \midrule
    % ======================= ROW 10 =======================
    \rowcolor{verylightgray}
    10
    & (0.005, 0.004)
    & \textbf{(0.000, 0.000)}
    & (0.013, 0.025)
    & (0.043, 0.041)
    & (17.809, 0.264)
    & (0.001, 0.000)
    & (0.002, 0.000)
    & (0.008, 0.004)
    \\
    % ======================= ROW 20 =======================
    20
    & (0.004, 0.003)
    & \textbf{(0.003, 0.005)}
    & (0.028, 0.030)
    & (0.066, 0.068)
    & (39.068, 0.437)
    & (0.518, 0.454)
    & (0.414, 0.216)
    & (1.770, 1.634)
    \\
    % ======================= ROW 40 =======================
    \rowcolor{verylightgray}
    40
    & \textbf{(0.009, 0.004)}
    & \textbf{(0.009, 0.004)}
    & (0.038, 0.037)
    & (0.112, 0.122)
    & (71.404, 0.459)
    & (4.711, 1.472)
    & (10.132, 2.430)
    & (60.077, 23.715)
    \\
    % ======================= ROW 50 =======================
    50
    & \textbf{(0.008, 0.004)}
    & (0.015, 0.014)
    & (0.667, 0.627)
    & (0.104, 0.013)
    & (85.426, 0.336)
    & (7.257, 2.137)
    & (14.928, 2.774)
    & (92.407, 22.611)
    \\
    % ======================= ROW 60 =======================
    \rowcolor{verylightgray}
    60
    & \textbf{(0.009, 0.004)}
    & (0.024, 0.025)
    & (4.986, 4.434)
    & (0.172, 0.100)
    & (99.726, 1.103)
    & (7.312, 1.775)
    & (18.441, 3.045)
    & (126.920, 31.016)
    \\
    % ======================= ROW 80 =======================
    80
    & \textbf{(0.008, 0.003)}
    & (0.394, 0.432)
    & (11.962, 9.007)
    & (0.337, 0.107)
    & (123.876, 0.738)
    & (10.319, 2.331)
    & (22.156, 5.513)
    & (144.083, 28.293)
    \\
    % ======================= ROW 90 =======================
    \rowcolor{verylightgray}
    90
    & \textbf{(0.008, 0.002)}
    & (2.433, 0.992)
    & (21.048, 9.212)
    & (0.311, 0.094)
    & (135.041, 0.945)
    & (13.988, 1.141)
    & (22.638, 6.413)
    & (159.869, 32.978)
    \\
    % ======================= ROW 100 =======================
    100
    & \textbf{(0.008, 0.003)}
    & (5.416, 1.397)
    & (469.161, 142.922)
    & (0.298, 0.035)
    & (145.419, 0.748)
    & (13.421, 3.041)
    & (29.211, 7.764)
    & (177.615, 25.508)
    \\
    % ======================= ROW 120 =======================
    \rowcolor{verylightgray}
    120
    & \textbf{(0.010, 0.002)}
    & (10.024, 4.780)
    & (560.380, 248.626)
    & (0.452, 0.110)
    & (162.739, 0.735)
    & (14.338, 6.238)
    & (34.206, 7.075)
    & (182.186, 35.903)
    \\
    % ======================= ROW 140 =======================
    140
    & \textbf{(0.010, 0.001)}
    & (15.925, 4.961)
    & (784.101, 273.030)
    & (0.440, 0.097)
    & (179.231, 0.664)
    & (17.916, 4.025)
    & (31.954, 8.984)
    & (249.601, 10.626)
    \\
    % ======================= ROW 150 =======================
    \rowcolor{verylightgray}
    150
    & \textbf{(0.010, 0.002)}
    & (20.127, 3.681)
    & (1054.898, 193.782)
    & (0.393, 0.027)
    & (185.540, 2.805)
    & (18.899, 4.439)
    & (38.051, 6.533)
    & (223.909, 43.492)
    \\
    % ======================= ROW 170 =======================
    170
    & \textbf{(0.010, 0.000)}
    & (24.700, 0.882)
    & (1033.337, 163.962)
    & (0.477, 0.124)
    & (197.188, 4.335)
    & (23.828, 2.837)
    & (42.323, 13.873)
    & (245.674, 39.698)
    \\
    % ======================= ROW 200 =======================
    \rowcolor{verylightgray}
    200
    & \textbf{(0.012, 0.000)}
    & (489.539, 64.347)
    & (942.219, 210.909)
    & (0.437, 0.143)
    & (208.397, 6.836)
    & (23.418, 7.994)
    & (45.841, 16.075)
    & (292.812, 29.301)
    \\
    % ======================= ROW 220 =======================
    220
    & \textbf{(0.014, 0.000)}
    & (503.676, 103.121)
    & (1156.416, 273.305)
    & (0.488, 0.158)
    & (203.871, 5.406)
    & (27.639, 10.269)
    & (48.729, 15.949)
    & (218.949, 84.314)
    \\
    % ======================= ROW 250 =======================
    \rowcolor{verylightgray}
    250
    & \textbf{(0.017, 0.000)}
    & (605.856, 45.683)
    & (979.074, 166.781)
    & (0.575, 0.036)
    & (200.591, 8.107)
    & (29.549, 9.451)
    & (63.678, 12.886)
    & (282.143, 24.587)
    \\
    \bottomrule
    \end{tabular}
    }
\end{table*}

\section{Experimental Setup}\label{sec4:experimentation}

In this section, we detail the experiments conducted to evaluate the three different approaches.

\subsection{Max-Cut Benchmark Description} \label{subsec:MaxCutBenchmark}
The experiments are carried out on a variety of graph instances of the Weighted Max-Cut problem (see Section \ref{subsec:MAXCUT}). Specifically, we evaluate each solver using a benchmark of 15 instances provided in \cite{eneko}. This selection enables a rigorous and meaningful comparison by leveraging a dataset with established reference results. The benchmark comprises 15 weighted graph instances of varying sizes, ranging from $N=10$ nodes to $N=250$ nodes, with edge weights uniformly distributed in the range $[0, 2]$. Table \ref{tab:maxcut_bench} summarizes a detailed description of all benchmark instances.

\begin{table}
\caption{Description of the Max-Cut instances. The first column refers to the number of nodes in the graph $N$ and the second to the number of edges.}
    \centering
    \begin{tabular}{c|c||c|c||c|c}
        \hline
         $N$ & Num. Edges & $N$ & Num. Edges & $N$ & Num. Edges \\
         \hline
         \hline
         10 & 37 & 80 & 2563 & 150 & 8822 \\
         20 & 148 & 90 & 3200 & 170 & 11486 \\
         40 & 570 & 100 & 3902  & 200 & 15916 \\
         50 & 1013 & 120 & 5696 & 220 & 19423 \\
         60 & 1450  & 140 & 7912 & 250 & 24831\\
         \hline
    \end{tabular} \label{tab:maxcut_bench}
\end{table}

\subsection{Evaluation Criteria} To gather sufficient data to draw reliable conclusions on each algorithm’s effectiveness and efficiency, we ran the experiment 10 times for each Max-Cut instance and solver. During each run, we recorded the minimum energy (or fitness value), execution time, and memory usage. To gauge solution quality, we relied on the averaged approximation ratio $\bar{\text{AR}}$, a measure indicating how closely a solver approximates the optimal solution. For a given instance, we compute it as: \begin{equation}\label{eq:averageAR} \bar{\text{AR}} = \frac{1}{R} \sum_{r=1}^R \frac{E_r^*}{E_g}, \end{equation} where $E_r^*$ is the minimum energy (or fitness value) of the $r$-th run, $R$ is the total number of runs, and $E_g$ is the ground state energy (i.e., the optimal solution) of the given instance. An $\bar{\text{AR}}$ approaching 1 indicates that the solver produces a near-optimal solution. Execution time and memory usage, like the approximation ratio, are averaged over all runs for each problem instance. Averaging the execution time allows us to evaluate each algorithm’s efficiency while tracking peak memory usage provides insight into resource demands. By gathering and comparing these metrics, we aim to provide a holistic perspective on the balance between solution quality, execution time, and memory usage for each method.

\section{Results}\label{sec5:results}

This section presents the findings from our experimental evaluation of the three algorithms --\(GAs\), \(GNN\), and \(DMRG\)-- on the Max-Cut instances under consideration. The analysis systematically examines the performance of each solver, focusing on the key metrics outlined in the previous section.

The experimental results are summarized in Tables~\ref{tab:energy_tab_reduced},~\ref{tab:time_in_minutes_reduced}, and~\ref{tab:memory_tab}, detailing the performance of each algorithm across three key metrics: averaged approximation ratio (see Eq.~\ref{eq:averageAR}), execution time, and memory usage, respectively. Each row in these tables corresponds to a specific Max-Cut instance, characterized by the number of nodes \(N\) in the graph, while the columns reflect the performance of the three solvers.

For the cGA, the last three columns report results for different population sizes: 500, 1000, and 2000 individuals. The results for the GA-OC and GNN are presented in the previous columns. The first three columns display the performance of the DMRG algorithm, evaluated under varying bond dimensions \(\chi\): \(2\), \(0.10N\), and \(0.20N\). All results are presented in the format (average value, standard deviation), and the best result for each instance (row) is highlighted in bold.

The results presented in Table~\ref{tab:energy_tab_reduced} show that the DMRG algorithm achieves superior performance across most Max-Cut instances, obtaining approximation ratios in the range \([0.96, 0.99]\), regardless of the instance size. This highlights the algorithm's scalability and robustness, with \(\chi=2\) yielding strong results, while larger bond dimensions provide marginal improvements. In contrast, the cGA excels on smaller instances, attaining near-optimal averaged approximation ratios. However, its performance diminishes as \(N\) increases, with approximation ratios approaching \(0.95\). Conversely, the GA-OC achieves better results as the number of nodes ($N$) increases. 

\begin{figure*}[t]
    \centering
    \includegraphics[width=0.6\linewidth]
    {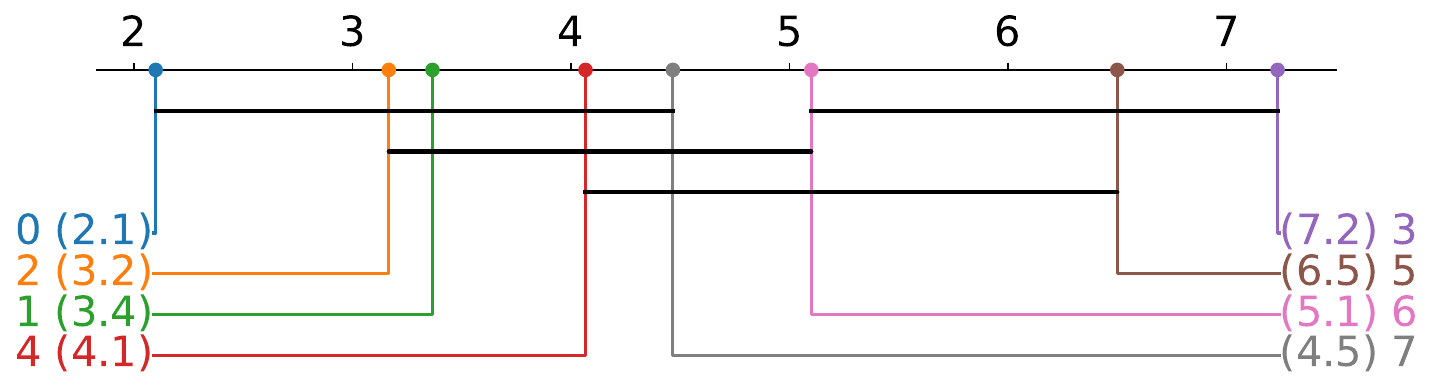}
    \caption{Critical Difference Diagram, with Friedman test $p\text{-value}<0.05$ and critical distance $CD=1.753$. The plot shows the average ranks (in brackets) achieved by each approach, calculated based on $\bar{\text{AR}}$. In this case, lower average ranks indicate better performance. The algorithms are indexed from 0 to 7 as follows: 
\textcolor{Col0}{\textbf{0} \texttt{DMRG 2}}, 
 \textcolor{Col1}{\textbf{1 }\texttt{DMRG $0.10N$}}, 
 \textcolor{Col2}{\textbf{2 }\texttt{DMRG $0.20N$}}, 
 \textcolor{Col3}{\textbf{3 } \texttt{GNN}}, 
 \textcolor{Col4}{\textbf{4 }\texttt{GA-OC}}, 
 \textcolor{Col5}{\textbf{5 } \texttt{cGA 500}}, 
 \textcolor{Col6}{\textbf{6 }\texttt{cGA 1000}}, 
\textcolor{Col7}{\textbf{7 } \texttt{cGA 2000}}.}
    \label{fig:cd-diagram}
\end{figure*}

To evaluate the statistical significance of the obtained results, the Friedman and the post-hoc Nemenyi tests are applied, following the guidelines in  \cite{demvsar2006statistical}. The former test determines whether different solvers are statistically different, while the pos-hoc test determines for which pair of strategies the statistical significance holds. In this case, the obtained Friedman statistic is $\chi^2_F=53.032$ with a $p\text{-value} = 3.65\times 10^{-9}$. Since this value is less than $0.05$, we can conclude that the differences among the different algorithms are statistically significant. The ranks produced by the Friedman test and the Nemenyi test reveal that top solvers are $\texttt{DMRG}$ with $\chi_{\text{bond}}=2, 0.1N, 0.20N$ and $\texttt{GA-OC}$ solvers, see Figure \ref{fig:cd-diagram} plot, where the average ranks (in $\bar{\text{AR}}$) achieved by each approach is shown. Note that lower average ranks are preferred as they perform better. The least effective algorithms are the \texttt{GNN}, \texttt{cGA 500}, and \texttt{cGA 1000}.

Regarding execution time, results shown in Table~\ref{tab:time_in_minutes_reduced} indicate that the DMRG approach with bond dimension \(\chi=2\) achieves the lowest execution times across most problem sizes, particularly for larger graphs. While increasing the bond dimension allows the MPS to capture more entanglement—or correlations among variables—it also leads to higher runtimes, especially for instances with \(N \geq 100\). Notably, as indicated by the averaged approximation results in Table~\ref{tab:energy_tab_reduced}, this additional computational cost is not always justified.

Although the execution time of the GNN does not surpass that of DMRG with \(\chi=2\), the model still exhibits acceptable runtimes. Conversely, the runtime of the GAs grows substantially with both population size \(P\) and problem size \(N\), resulting in the longest execution times among all tested methods—particularly when \(P=2000\). However, considering the population of GA-OC, we show that among the GAs, GA-OC is the worst.

Memory consumption, reported in Table~\ref{tab:memory_tab}, reveals that DMRG is the most demanding method in terms of memory usage, especially when the bond dimension increases beyond \(2\) and the graph size reaches \(N\geq100\). Although \(\chi=2\) is relatively more efficient than higher bond dimensions, it still consumes more memory than both the GNN and GA approaches. In contrast, the GNN maintains a modest memory footprint (often under 1\,MiB even for large \(N\)), providing a significant advantage in resource-constrained environments. Although memory usage for the GAs increases with population size, it remains considerably lower than that of the high-bond-dimension DMRG.

\begin{table*}[htbp]
    \centering
    \caption{Memory usage (in MiB) for DMRG, GNN, GA-OC, and cGA. 
    The best (lowest) average memory for each $N$ is in bold.}
    \label{tab:memory_tab}
    \resizebox{\textwidth}{!}{%
    \begin{tabular}{c||ccc|c|c|ccc}
    \toprule
    \multirow{2}{*}{$N$}
      & \multicolumn{3}{c|}{DMRG}
      & \multirow{2}{*}{GNN}
      & \multirow{2}{*}{GA-OC}
      & \multicolumn{3}{c}{cGA} \\
    & $\chi_{\text{bond}}=2$
      & $\chi_{\text{bond}}=0.10N$
      & $\chi_{\text{bond}}=0.20N$
      &
      & %\texttt{Pop.Size}=300
      & \texttt{Pop.Size}=500
      & \texttt{Pop.Size}=1000
      & \texttt{Pop.Size}=2000 \\
    \midrule
    \rowcolor{verylightgray}
    10
      & (4.425, 0.000)
      & (4.426, 0.000)
      & (4.426, 0.000)
      & (0.476, 1.244)
      & \textbf{(0.347, 0.000)}
      & (0.409, 0.000)
      & (0.689, 0.000)
      & (1.246, 0.000)
      \\
    20
      & (4.508, 0.000)
      & (4.510, 0.000)
      & (4.518, 0.000)
      & \textbf{(0.075, 0.067)}
      & (0.471, 0.000)
      & (0.609, 0.000)
      & (1.089, 0.000)
      & (2.046, 0.000)
      \\
    \rowcolor{verylightgray}
    40
      & (5.078, 0.000)
      & (5.104, 0.000)
      & (5.192, 0.000)
      & \textbf{(0.117, 0.115)}
      & (0.713, 0.001)
      & (1.009, 0.000)
      & (1.889, 0.000)
      & (3.646, 0.000)
      \\
    50
      & (5.726, 0.000)
      & (5.774, 0.000)
      & (5.931, 0.000)
      & \textbf{(0.108, 0.013)}
      & (0.832, 0.001)
      & (1.209, 0.000)
      & (2.289, 0.000)
      & (4.446, 0.000)
      \\
    \rowcolor{verylightgray}
    60
      & (6.727, 0.000)
      & (6.814, 0.000)
      & (7.048, 0.000)
      & \textbf{(0.195, 0.116)}
      & (0.954, 0.002)
      & (1.409, 0.000)
      & (2.689, 0.000)
      & (5.246, 0.000)
      \\
    80
      & (9.840, 0.000)
      & (10.051, 0.000)
      & (18.519, 0.000)
      & \textbf{(0.316, 0.097)}
      & (1.193, 0.001)
      & (1.809, 0.000)
      & (3.489, 0.000)
      & (6.846, 0.000)
      \\
    \rowcolor{verylightgray}
    90
      & (12.023, 0.083)
      & (12.316, 0.084)
      & (28.411, 0.078)
      & \textbf{(0.283, 0.082)}
      & (1.315, 0.002)
      & (1.974, 0.035)
      & (3.854, 0.035)
      & (7.646, 0.000)
      \\
    100
      & (14.789, 0.007)
      & (16.032, 0.003)
      & (41.913, 0.003)
      & \textbf{(0.272, 0.032)}
      & (1.435, 0.002)
      & (2.139, 0.000)
      & (4.219, 0.000)
      & (8.446, 0.000)
      \\
    \rowcolor{verylightgray}
    120
      & (22.431, 0.009)
      & (29.165, 0.003)
      & (83.108, 0.004)
      & \textbf{(0.387, 0.088)}
      & (1.675, 0.003)
      & (2.539, 0.000)
      & (5.019, 0.000)
      & (10.046, 0.000)
      \\
    140
      & (33.115, 0.007)
      & (48.771, 0.003)
      & (149.058, 0.003)
      & \textbf{(0.374, 0.077)}
      & (1.915, 0.003)
      & (2.939, 0.000)
      & (5.819, 0.000)
      & (11.646, 0.000)
      \\
    \rowcolor{verylightgray}
    150
      & (39.776, 0.007)
      & (61.561, 0.004)
      & (193.891, 0.002)
      & \textbf{(0.331, 0.019)}
      & (2.034, 0.002)
      & (3.139, 0.000)
      & (6.219, 0.000)
      & (12.446, 0.000)
      \\
    170
      & (55.987, 0.008)
      & (94.106, 0.003)
      & (312.792, 0.004)
      & \textbf{(0.390, 0.097)}
      & (2.278, 0.003)
      & (3.539, 0.000)
      & (7.019, 0.000)
      & (14.046, 0.000)
      \\
    \rowcolor{verylightgray}
    200
      & (88.615, 0.000)
      & (164.314, 0.003)
      & (583.931, 0.002)
      & \textbf{(0.392, 0.146)}
      & (2.615, 0.042)
      & (4.139, 0.000)
      & (8.219, 0.000)
      & (16.446, 0.000)
      \\
    220
      & (116.628, 0.000)
      & (228.572, 0.003)
      & (843.623, 0.002)
      & \textbf{(0.397, 0.124)}
      & (2.774, 0.003)
      & (4.539, 0.000)
      & (9.018, 0.000)
      & (17.976, 0.000)
      \\
    \rowcolor{verylightgray}
    250
      & (169.426, 0.017)
      & (358.646, 0.000)
      & (1384.402, 0.003)
      & \textbf{(0.447, 0.027)}
      & (3.134, 0.006)
      & (5.139, 0.000)
      & (10.218, 0.000)
      & (20.376, 0.000)
      \\
    \bottomrule
    \end{tabular}
    }
\end{table*}

\section{Conclusion and Future Work}
\label{sec6:conclusions}

In this study, we employ eight distinct algorithms, categorized under three perspectives- Metaheuristic, Deep Learning, and Quantum-inspired -- to address the Weighted Max-Cut problem. Specifically, we introduce a scalable MPO representation of the Weighted Max-Cut Hamiltonian, facilitating efficient handling of large graph instances. This paper evaluates three representative algorithms applied to 15 weighted graphs: GAs representing metaheuristic models, a GNN as a deep learning approach, and DMRG as a tensor network-based technique.

Based on the results obtained, the following conclusions can be drawn:

\begin{itemize}
    \item The top solvers, in terms of solution quality, are $\texttt{DMRG}$ with $\chi_{\text{bond}}=2, 0.1N, 0.20N$ and the $\texttt{GA-OC}$ solvers.
    \item The DMRG with a bond dimension of \(\chi=2\) emerges as the most balanced and time-efficient method for large Max-Cut instances, maintaining high solution quality with relatively short runtimes, albeit at the expense of higher memory usage.
    \item GA-OC is the best algorithm among the GAs in terms of solution quality, but it performs worse than cGA in terms of time. It is also more efficient in memory usage due to its smaller population size.
    \item The cGAs perform optimally for smaller instances, delivering the best solutions; nevertheless, it requires significantly more time and a larger population size to remain competitive on larger problems.
    \item The GNN approach excels in minimizing memory consumption; however, its solution quality and runtime exhibit less consistency, particularly for larger graphs.
    \item Consequently, the selection among these methods hinges primarily on the trade-offs between runtime, memory availability, and the requisite solution quality.
\end{itemize}

Despite its strong performance, DMRG presents a notable limitation. Although the objective function defined in Equation \eqref{eq:max-cut1} can be mapped to an MPO, this process is typically challenging. Even minor alterations in the problem Hamiltonian necessitate reformulating the automata to derive a new MPO representation, thereby increasing the method's complexity. This issue does not arise with the GA and GNN approaches. Furthermore, it is critical to acknowledge that a bond dimension of \(\chi=2\) may be insufficient to capture the entanglement in certain problem instances. In such cases, the computational cost can escalate considerably, rendering classical models a more viable alternative.

As future research directions, we propose to extend this work in several key areas. We aim to investigate a broader array of combinatorial problems, including those with additional constraints such as the Traveling Salesman Problem (TSP), and to integrate additional algorithms (classical and quantum-inspired) to facilitate a more thorough performance. We also intend to develop a model-agnostic MPO generator capable of automatically constructing representations for any Hamiltonian instance, thus obviating the need for manual automata design and enhancing the overall workflow. Additionally, we plan to explore parallelization strategies for the Genetic Algorithm, including hyperparameter tuning, to improve its scalability and efficiency. We also aim to refine the basic DMRG implementation used in this study by incorporating state-of-the-art techniques. Moreover, we intend to experiment with more sophisticated GNN architectures, given that the current GNN implementation is relatively rudimentary, and to assess the performance of all three methods on GPU architectures, as our present experiments were conducted on CPUs. These extensions are anticipated to refine each approach and yield deeper insights into the trade-offs and synergies among classical, deep learning, and quantum-inspired optimization methods.

\section*{Code Availability}
The source code used to implement the methods and reproduce the results presented in this paper will be publicly available upon manuscript acceptance.

\section*{Acknowledgments}

This work has been supported by the Basque Government through the ELKARTEK Program (KUBIT project ref. KK-2024/00105) and the Basque Government through Plan complementario comunicaci\'on cu\'antica (EXP. 2022/01341) (A/20220551). Aitor Morais acknowledges the partial funding of his doctoral research at the University of Deusto, within the D4K (Deusto for Knowledge) team on applied artificial intelligence and quantum computing technologies. During the preparation of this work, the authors used Microsoft Copilot to improve the language and readability of the manuscript. After using this tool/service, the authors have reviewed and edited the content as needed and take full responsibility for the content of the publication.

% \printbibliography % Prints the bibliography
\bibliography{sample-base}

\end{document}